\documentclass[intlimits,twoside,a4paper]{article}

\usepackage{amsmath,amssymb}
\usepackage{graphicx}

\usepackage[cp1251]{inputenc}
\usepackage[eqsecnum]{cmpj3}

\issue{2026}{29}{1}{13602}
\doinumber{10.5488/CMP.29.13602}
\title[Symmetric mixtures in pores]%
{Symmetric mixtures in slit-like pores with selective walls}%
\author[A. Patrykiejew]{A. Patrykiejew\orcid{0000-0001-6586-3606}}
\address{Department of Theoretical Chemistry, Institute of Chemical Sciences, Faculty of Chemistry, Maria Curie-Sklodowska University, 20031 Lublin, Poland}
\date{Received 27 October 2025; revised 01 January 2026; accepted 11 January 2026; published 30 March 2026}

\begin{document}

\maketitle

\begin{abstract}
Symmetric mixtures characterized by high negative geometric and energetic non-additivity do not exhibit 
phase separation in the bulk. However, the phase separation occurs when such mixtures are confined 
in slit pores with selective walls. It is demonstrated that the wall selectivity affects 
 the pore filling. When the difference of the interaction energies between the mixture components and pore walls is lower than a certain
 threshold value, condensation occurs between a dilute phase and the mixed liquid. When this difference exceeds the threshold value, the pore filling 
 may occur in two steps. The first is the condensation of a dilute phase into the demixed liquid, and the second step leads to the formation of the 
 mixed liquid. We have elucidated the changes in the phase behavior caused by  non-additivity of symmetric mixtures, 
 and by the difference in the interaction energies of the components with pore walls.

\end{abstract}

\section{Introduction}
\label{INTRODUCTION}

The model of symmetric mixture (SM), although very simple, is characterized by rich physics~\cite{WSN,MPS2010,APSS2010,almarza,MPR2013,AP-PRE,vlot-liq,vlot1,vlot2}. 
In such mixtures, the components, A and B, have the same properties and  the interactions between the AA and BB pairs are identical, while 
the interaction between the AB pairs is different due to 
the energetic and/or geometric non-additivity. The interaction energy between a pair of particles, $i$ and $j$, is characterized by the depth, 
$\varepsilon_{ij}$, and the range, $\sigma_{ij}$, of the potential well. In the case of a symmetric mixture, $\varepsilon_{AA}=\varepsilon_{BB}=\varepsilon$ 
and, $\sigma_{AA}=\sigma_{BB}=\sigma$, while $\varepsilon_{AB}$ and 
$\sigma_{AB}$ are different, and may be defined as $\varepsilon_{AB}=e\varepsilon$ and $\sigma_{AB}=s\sigma$. Thus, the  parameters $e$ and $s$ describe the energetic
and geometric non-additivity, respectively.

When the symmetric mixture exhibits only energetic non-additivity, i.e., $e<1$, and $s=1$, the bulk mixtures were found to 
exhibit a demixing transition, which may occur in the solid and/or liquid
phases~\cite{WSN,MPS2010,tavares,ref1,ref2,ref3}. When demixing occurs in the liquid phase, the transition may be discontinuous (first-order) or continuous (second-order) 
depending on the value of $e$ and the temperature. 
In general, the tendency towards phase separation is enhanced when the parameter $e$ becomes lower. 

On the other hand, when the mixture shows only geometric non-additivity, i.e.,  $s\ne 1$
and $e=1$, only the packing effects affect the properties of the mixture, especially in the high-density 
region. In particular, it was demonstrated~\cite{vlot2} that solid  phases are mixed, or partially mixed, and form crystals of different structures 
depending on $s$.
In the liquid phase, the AA and BB pairs are more likely to appear when $s>1$, and this leads to the formation of clusters consisting of like particles, though the densities of both components remain the same.
On the other hand, when $s<1$, the formation of AB pairs is favored, which enhances miscibility.
      
When the parameters, $e$ and $s$, are both lower than unity, the situation becomes more complex, since low values of $e$ favor demixing, while low values of $s$ 
favor mixing~\cite{AP-PRE,AP-JML}. 
Competing energetic and geometric non-additivity causes the demixing transition in the liquid to occur only when 
$s$ is larger than a certain threshold value $s_o(e)$.
 When $s$ is lower than $s_o(e)$, phase separation is suppressed completely, and the phase diagrams of such mixtures 
become qualitatively the same as in single-component systems~\cite{AP-PRE,AP-PCCP}. Thus, the vapor condenses into 
the mixed liquid, and the vapor-liquid transition terminates in the critical point. In such systems, the solid phase is also mixed. 

The behavior of SMs adsorbed in slit-like pores has been studied theoretically, using the lattice and off-lattice 
density functional theory~\cite{POR3,POR4,POR5,POR6}, and computer simulations~\cite{POR7,Sholl,AP-JML}.  
The primary objective of those studies was to determine how geometric constraints and the interaction 
of fluid particles with the pore walls  affect the phase separation in liquids. It should be noted that the vast 
majority of those studies were devoted to mixtures in which $s=1$, and only the parameter $e$ was less than unity.   

The effects of confinement on demixing transition in mixtures with $e<1$ and $s\ne1$ have also been 
considered~\cite{AP-JML,POR7}, but mostly in systems with non-selective walls, i.e., with the same 
energies of interaction between the mixture components and the pore walls. 
The symmetric mixtures with $s>1$ in slit-like pores with crystalline walls of square symmetry were studied by Monte Carlo simulation~\cite{POR7}, 
and shown to exhibit the formation of condensed phases with different sequences of demixed layers, despite practically the 
same total densities of both components in the system.

Recently, we have studied the behavior of symmetric mixtures with $e=0.6$, and $s>s_o(e)$, in slit pores with non-selective walls~\cite{AP-JML}.  
It has been shown that geometrical constraints are of rather limited importance and lead to qualitative changes in the phase behavior
 only in very narrow pores. In such cases, phase separation does not occur at all, 
and the gas condenses into the mixed liquid up to 
the critical temperature of capillary condensation. In wider pores, the phase behavior was show to be qualitatively the same as in bulk systems
with $s>s_o(e)$.

An interesting, and as yet not studied problem, is the effect of geometric confinement on the behavior of highly non-additive symmetric 
mixtures, with $s<s_o(e)$, which do not exhibit a demixing transition in the bulk, and when the components interact differently with the pore walls. 
Such a situation may correspond to racemic mixtures placed in pores with chiral walls~\cite{Xue}.
Our goal here is to determine whether the wall selectivity can lead to the phase separation in the confined liquid, and how this affects the phase behavior of systems 
characterized by different magnitudes of the parameters $e$ and $s$, different pore widths, as well as different energies of interaction of the mixture components with the walls.  

The paper is organized as follows. The next section presents the model and the simulation method used.  
In section III, we present and discuss the results. 
The paper concludes in section IV, which contains a brief summary of our findings and final remarks.

\section{The model and simulation methods}
\label{sec-2}

We consider mixtures, in which
the pair of particles, $i$ and $j$, interact via the (12,6) Lennard-Jones potential
\begin{equation}
u_{ij}(r) = \left\{ \begin{array}{ll}
4\varepsilon_{ij}\left[(\sigma_{ij}/r)^{12} - (\sigma_{ij}/r)^{6}\right],   & r \leqslant r_{\rm max}, \\
0,                                                & r > r_{\rm max}.
\end{array}
\right.
\label{eq:01a}
\end{equation}
In a symmetric mixture,
$\sigma_{AA}=\sigma_{BB}=\sigma$ and $\varepsilon_{AA}=\varepsilon_{BB}=\varepsilon$, while
$\sigma_{AB}$ and $\varepsilon_{AB}$ are different and given by $\sigma_{AB}=s\sigma$ and $\varepsilon_{AB} = e\varepsilon$.
The potential is cut at the distance $r_{\rm max}=3\sigma$, and
$\sigma$ and $\varepsilon$ are taken as the units of length and energy, respectively.

The interaction of particles with the
pore walls is modelled with the potential
\begin{equation}
v_i(z) = \varepsilon_{gs,i}\left[\left(\frac{\sigma}{z}\right)^9-\left(\frac{\sigma}{z}\right)^3+
\left(\frac{\sigma}{H-z}\right)^9-\left(\frac{\sigma}{H-z}\right)^3\right],\quad \mbox{with}\,\, i=A\, \mbox{or}\, B,
\label{eq:01c}
\end{equation}
where $H$ is the pore width.

In this work, we considered the systems with different $\varepsilon_{gs,A}$ and $\varepsilon_{gs,B}$, and
assumed that $\varepsilon_{gs,A} > \varepsilon_{gs,B}$, while the
average value, $\bar{\varepsilon}_{gs} = 0.5(\varepsilon_{gs,A} + \varepsilon_{gs,B})$, is fixed.
By changing $\Delta V = \varepsilon_{gs,A}-\varepsilon_{gs,B}$, we studied how the wall
selectivity affects the behavior of mixtures, with respect to non-selective walls ($\Delta V=0$).
Three different values of $\bar{\varepsilon}_{gs}$ equal to 5, 10, and 15 are used.

In order to study the phase behavior, we used the Monte Carlo method in the grand canonical ensemble (GCMC), as described in our earlier
paper~\cite{AP-JML}. The simulations are carried out in cells of the size $30\times 30\times H$, with standard periodic boundary
conditions applied in the directions parallel to the pore walls. The attempted moves within a single Monte Carlo step involved the creation or annihilation of particles,
as well as translation and identity exchange
of randomly chosen particles. The number of Monte Carlo steps, necessary to equilibrate the system and to calculate the averages, is taken to be between $(10^7 - 10^8)N$, where  $N$ is the momentary number of particles in the system, depending on the temperature. Ten times longer runs were needed at temperatures close to the critical points.

We recorded
the average potential energy (per particle), $\langle u^{\ast}\rangle$, the densities of both components $\rho_k$ ($k=$ A or B), and the density profiles $\rho_k(z)$.
Besides, we calculated the order parameter
\begin{equation}
m = \frac{|\rho_A-\rho_B|}{\rho_A+\rho_B},
\end{equation}
and the order parameter profiles
\begin{equation}
m(z) = \frac{|\rho_A(z)-\rho_B(z)|}{\rho_A(z)+\rho_B(z)}.
\end{equation}
These parameters allow the determination of  differences in total densities and local densities of the mixture components.

The calculated adsorption-desorption isotherm were used to estimate the locations of phase transitions.

To obtain a precise information about the locations of discontinuous (first-order) phase transitions, one should determine
free energies of the coexisting phases~\cite{LB00}. In principle, it is possible when the system can be transformed into a reference state of known
free energy. Although a method to calculate the absolute free energy of one-component disordered systems was proposed~\cite{TSCH}, it cannot be directly
applied to mixtures. Another possible way to obtain the information about free energies of coexisting phases is the use of histogram reweighting methods~\cite{LB00,NW1}.

However, our aim here is to elucidate the effects of the wall selectivity on the phase behavior qualitatively only; hence, we have not used those time-consuming methods and relied  on the results
stemming from the recorded quantities.

Since the results at low temperatures are affected  by the presence of long-living metastable states, leading to quite broad hysteresis loops on the adsorption and desorption branches of isotherms, we were able to obtain reliable results at rather high temperatures.
Nevertheless, we were able to estimate the phase diagrams for several systems and to draw conclusions concerning the influence of the wall selectivity on the phase behavior.

\section{Results and discussion}
\label{sec-3}

\begin{figure}[h]
\begin{center}
\includegraphics[scale=0.38]{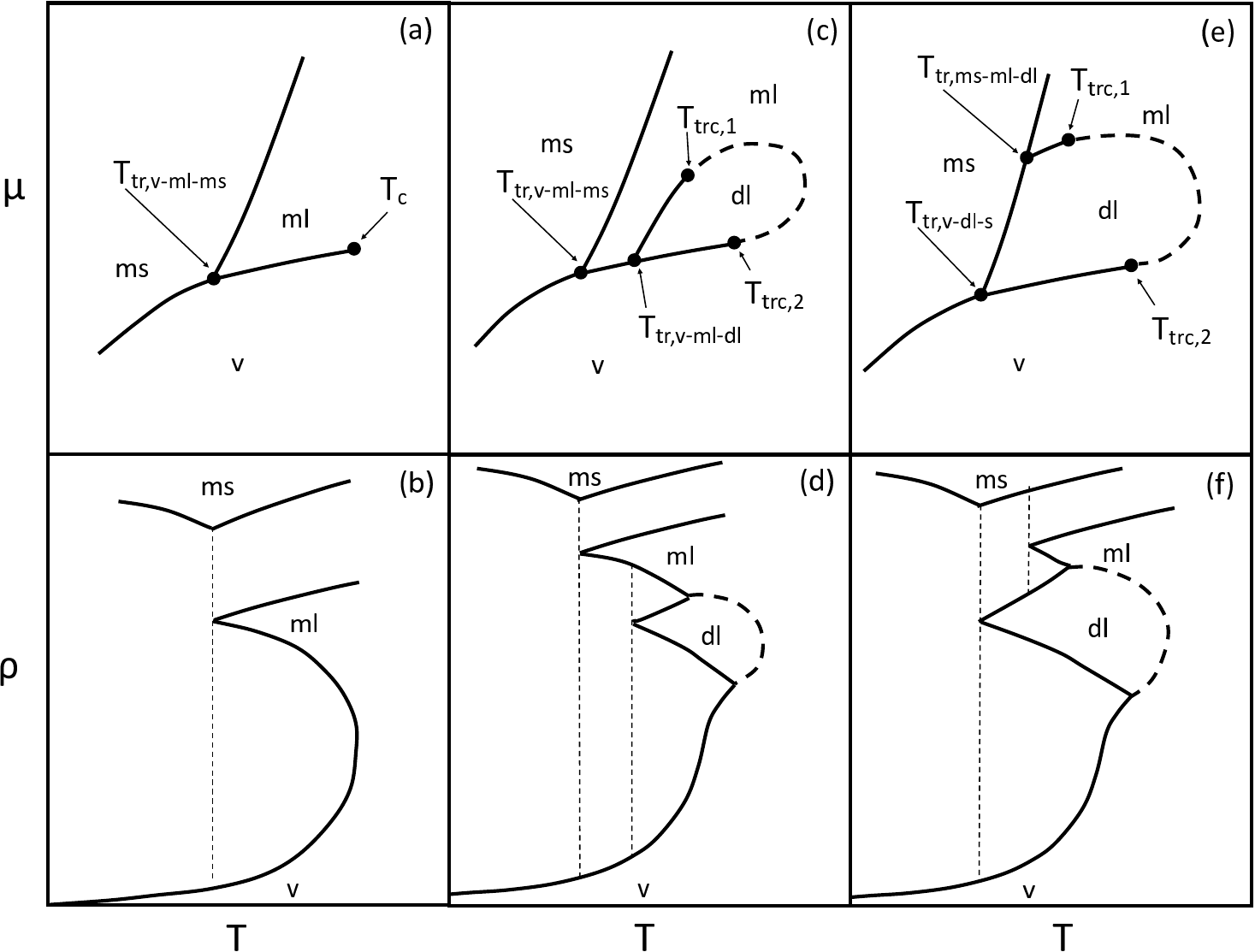}
 \caption{The schematic representations of the bulk phase diagram for the mixtures  with high negative energetic and geometric non-uniformity. Panels a, c, and e (b, d, and f)
 show the $\mu-T$ ($\rho-T$) projections of the phase diagrams for $s<s_o(e)$, $s\in (s_o(e),s_1(e))$, and $s>s_1(e)$, respectively.} 
 \label{fig_1}
\end{center}
 \end{figure}

We begin by recalling the basic information about the phase behavior of bulk symmetric mixtures
 with negative energetic and geometric non-additivity, i.e., when the parameters $e$ and $s$ are both lower than unity.
In general, low values of $e$ favor demixing, while  
low values of $s$ enhance mixing.  
In particular, when the parameters, $e$ and $s$, are considerably lower than unity, a competition between mixing and demixing 
leads to rather peculiar changes in the phase behavior of symmetric mixtures~\cite{AP-PRE,AP-PCCP}.   
In particular, when the parameter $s$ is lower than a certain value, $s_o(e)$, the liquid does not exhibit demixing transition at all, 
and phase diagrams of such systems are qualitatively the same as in one-component systems (see panels a and b of figure~\ref{fig_1}).
As soon as the parameter $s$ becomes larger than $s_o(e)$, the topology of phase diagrams change, since the liquid phase exhibits 
phase separation over a certain range of temperatures above the triple point, $T_{\textrm{tr,v-dl-ml}}$, in which the vapor coexists with the mixed and demixed liquids, up
to the tricritical point, $T_{\textrm{trc,2}}$, which replaces 
the critical point of the vapor-liquid transition. In  $T_{\textrm{trc,2}}$, the coexisting phases (vapor, mixed liquid, and demixed liquid) all become critical.
The tricritical point, $T_{\textrm{trc,2}}$, is the onset of the 
continuous demixing transition, which occurs along the so-called $\lambda$-line and forms a closed loop, terminating in the tricritical 
point $T_{\textrm{trc,1}}$ (see panels c and d of figure~\ref{fig_1}).
At temperatures below $T_{\textrm{tr,v-dl-ml}}$, the vapor condenses into the mixed liquid, and this transition occurs at temperatures down to the 
triple point $T_{\textrm{tr,v-ml-ms}}$, in which the vapor coexists with the mixed liquid and solid phases. 
When the parameter $s$ is gradually increased, the triple point $T_{\textrm{tr,v-dl-ml}}$ moves down towards the triple point $T_{\textrm{tr,v-ml-ms}}$. 
Ultimately, when $s$ reaches the value $s_1(e)$, these two triple points merge
into the quadruple point, $T_{\textrm{q,v-dl-ml-ms}}$, in which the vapor coexists with the mixed and demixed liquids, and the mixed solid.
For $s>s_1(e)$, the vapor always condenses into the demixed liquid, which undergoes transition into the mixed liquid at sufficiently high densities. 
In such cases, the loop of demixing transition terminates in the triple point $T_{\textrm{tr,ms-ml-dl}}$, located along the liquid-solid coexistence (see panels e and f of figure~\ref{fig_1}).

\begin{figure}[h!] 
	\begin{center}
		\includegraphics[scale=0.4]{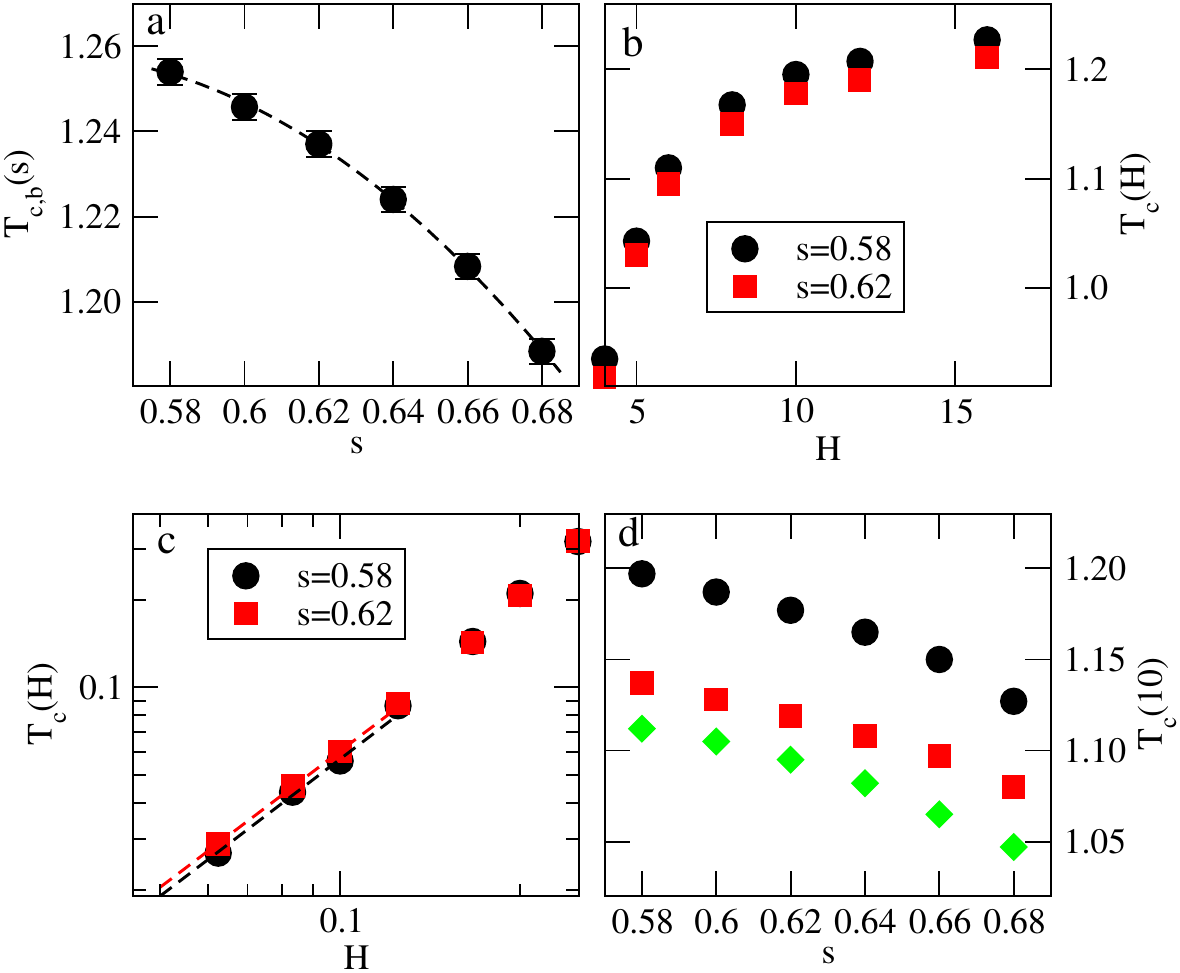}
	\end{center}
	\caption{(Colour online) (Panel a) Changes of the bulk critical temperature, $T_{c,b}$, with $s$. (Panel b) Changes of the capillary condensation critical temperature, $T_c(H)$, 
		with pore width, obtained for two different values of $s$, are given in the figure. (Panel c) The log-log plots of $\Delta T_c =T_{c,b}-T_c(H)$ against $1/H$ for the 
		same systems as shown in panel b. Dashed lines are the fits to the scaling relation (\ref{eq-scala}). (Panel d) Changes of the capillary condensation critical 
		temperature with $s$, for the pore of $H=10$, and different values of $\bar{\varepsilon}_{gs}$ equal to 5 (filled circles), 10 (filled squares), and 15 (filled diamonds). 
		All results correspond to the systems with $e=0.6$.}
	\label{fig_2}
\end{figure}

In this work, we consider the systems with $s<s_o(e)$, which do not exhibit demixing transition in the bulk.
The first series of results have been obtained for mixtures characterized by $e=0.6$, and $s$ ranging between 0.58 and 0.68. 
Our earlier studies~\cite{AP-PRE,AP-JML}  showed that $s_{o}(0.6)$ is equal to 0.695. The  
estimated changes of the bulk critical temperature, $T_{c,b}$, with $s$, are presented in figure~\ref{fig_2}a.
One should note a gradual decrease in $T_{c,b}$ with increasing $s$. This can be easily understood by taking into account that the 
stability of the mixed liquid phase decreases when the parameter $s$ approaches $s_o(e)$. 
One also expects a lack of demixing when such mixtures are adsorbed in slit pores with non-selective walls, i.e, when $\Delta V=0$. 
In this case, capillary condensation occurs between the gas and the mixed liquid, and this transition terminates in the critical temperature, $T_c(H)$. 
Of course, $T_c(H)$ increases with $H$, and is expected to converge to the bulk critical temperature when $H$ tends to infinity. 
It is known that in one-component fluids confined between the walls of slit pores, the following scaling relation is fulfilled~\cite{Fish01}
\begin{equation}
 T_{c,b}-T_c(H)\propto H^{-1/\nu}\;\;, 
 \label{eq-scala}
\end{equation}
with $\nu$ being the bulk correlation length exponent. 
Panel b of figure~\ref{fig_2} shows the changes of $T_c(H)$ for the systems with $s=0.58$ and 0.62, and 
with $\bar{\varepsilon}_{gs}=5$.
Although we have studied only rather narrow pores, with $H$ up to 16, 
and our estimations of critical temperatures are only approximate, nonetheless, the fit of data for $H\geqslant 8$ to the above equation (\ref{eq-scala})
 gives $\nu\approx 0.64$ (see panel c of figure~\ref{fig_2}). This value is very close to $\nu\approx 0.63$, corresponding to the universality class of 
3D Ising model~\cite{3D-Ising}, which also includes the mixed liquid phase considered. 
It is expected that for a fixed $H$, the critical temperature of capillary condensation also decreases with an increase in $s$. 
This has already been demonstrated in panel b of figure~\ref{fig_2}, which shows that $T_c(H)$  
for the system with $s=0.62$  is lower than in the case of $s=0.58$, independently of $H$.
Panel d of figure~\ref{fig_2} shows the changes in the critical temperature of capillary condensation with increasing $s$ in the pore of $H=10$, 
and for different magnitudes of the fluid interaction energy with the pore walls. 

\begin{figure}[h!]
	\begin{center}
		\includegraphics[scale=0.51]{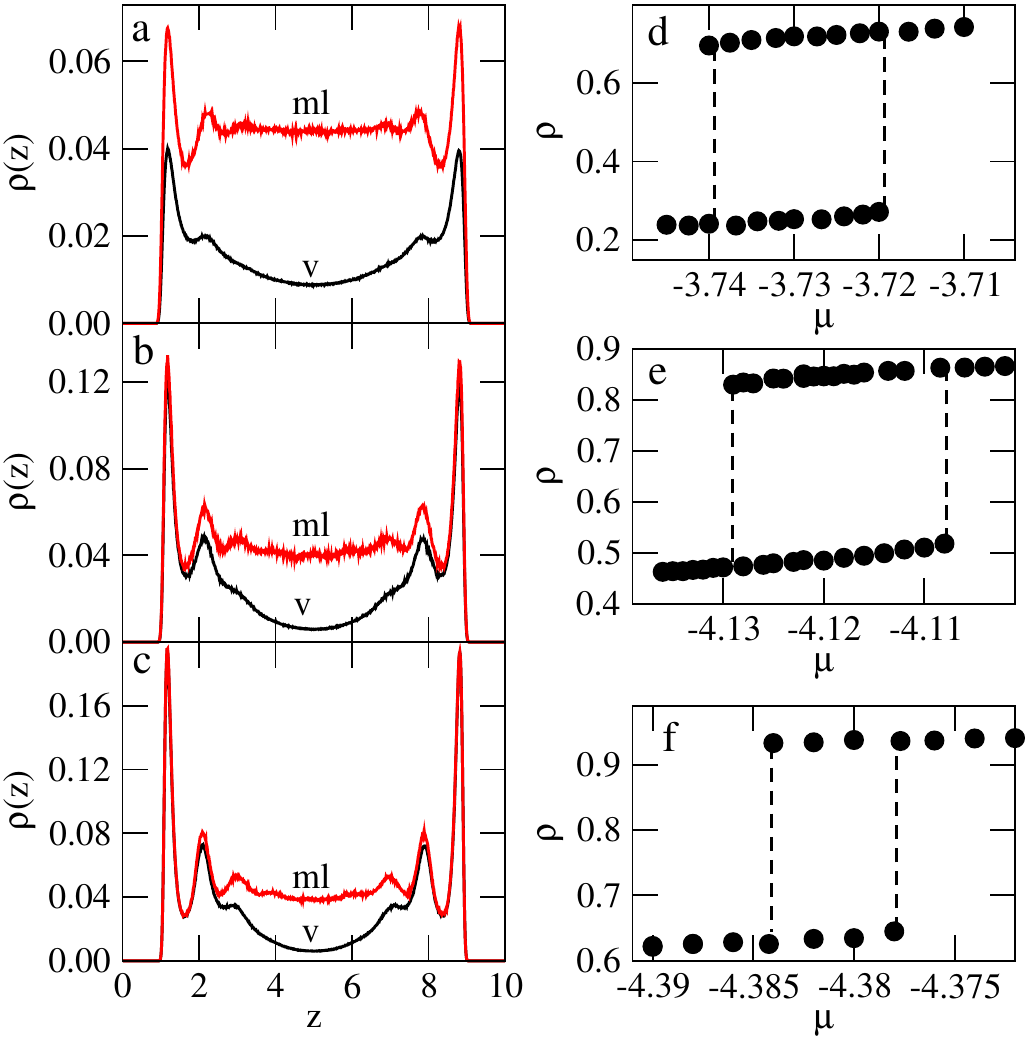}
	\end{center}
	\caption{(Colour online) Panel a--c show the density profiles recorded for the systems with $e=0.6$ and $s=0.68$, for $\bar{\varepsilon}_{gs}=5$, $T=1.10$ and $\mu=-3.73$ (panel a), 
		$\bar{\varepsilon}_{gs}=10$, $T=1.05$ and $\mu=-4.12$ (panel b), and $\bar{\varepsilon}_{gs}=15$, $T=1.02$ and $\mu=-4.38$ (panel c). Panels d-f show the 
		adsorption-desorption isotherms for the same systems as in panels a--c.} 
	\label{fig_3}
\end{figure}

The above results demonstrate that
apart from a gradual decrease of $T_c(10)$ with $s$,
 the  critical temperature of capillary condensation decays when the fluid-wall interaction becomes stronger. 
An increase of $\bar{\varepsilon}_{gs}$ leads to a gradual widening of the dense, and structured, adsorbed layers adjacent to the walls, hence the confined
fluid gains higher density and thickness prior to the capillary condensation. Thus, condensation occurs in a smaller central part of the 
pore when $\bar{\varepsilon}_{gs}$ becomes higher. Panels a, b, and c of figure~\ref{fig_3} present the 
examples of density profiles, recorded for the systems with the fixed $H=10$ and $s=0.68$, but with different values of $\bar{\varepsilon}_{gs}$, 
at the temperatures equal to about $0.97T_c(10)$. These profiles show that when $\bar{\varepsilon}_{gs}$ increases, the layered film close 
to the pore walls becomes thicker and reaches higher densities. Besides, the density of surface films is less affected by the capillary condensation when $\bar{\varepsilon}_{gs}$ is high. 
In the case of $\bar{\varepsilon}_{gs}=5$, the density of the layers adjacent to the pore walls 
rapidly increases when the capillary condensation takes place, while in the system with $\bar{\varepsilon}_{gs}=15$, the densities of the two layers at each wall are practically the same in the dilute and condensed phases.
On the other hand, the condensed 
mixed liquid in the central part of the pore has a density nearly independent of $\bar{\varepsilon}_{gs}$. 
Nonetheless, the total density of condensed liquid increases with 
$\bar{\varepsilon}_{gs}$, as it is illustrated by the isotherms given in panels d, e, and~f of figure~\ref{fig_3}.

\begin{figure}[h]
	\begin{center}
		\includegraphics[scale=0.5]{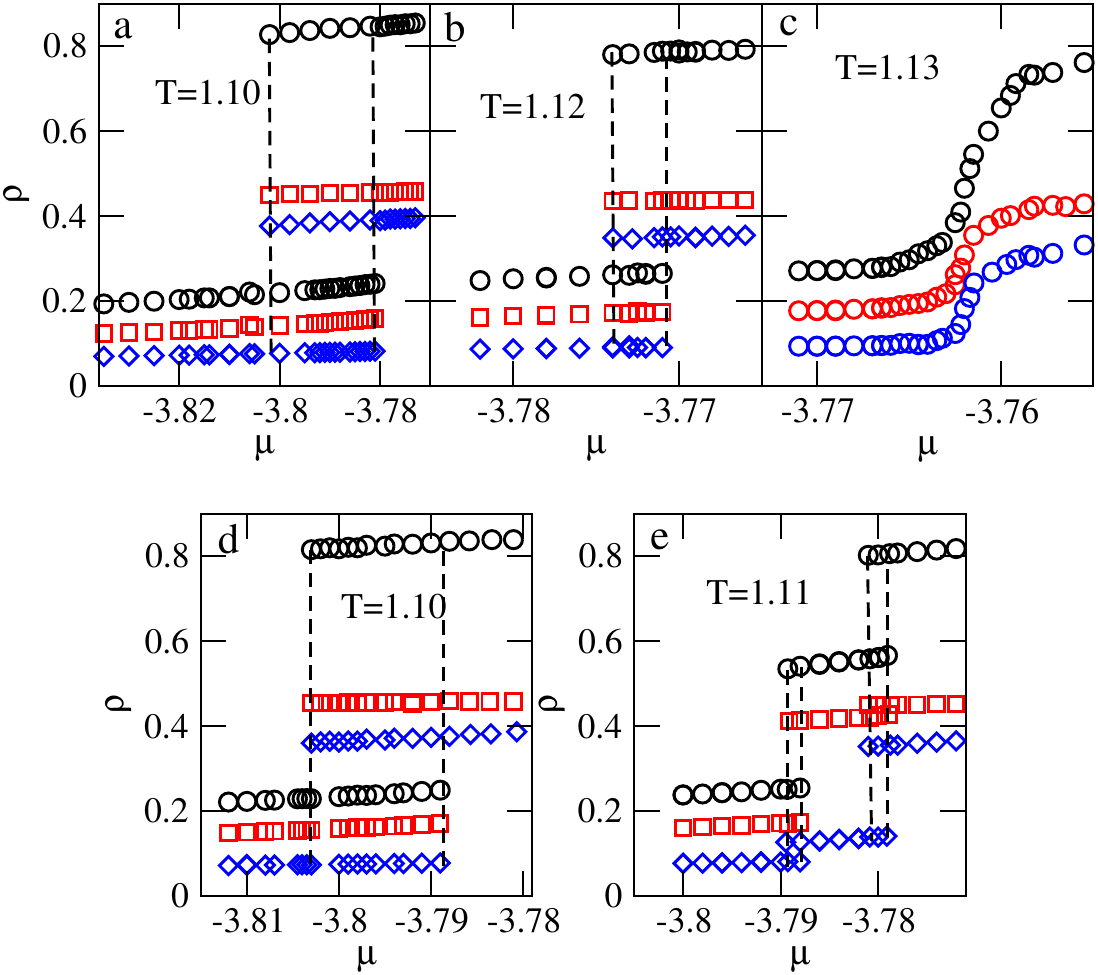}
	\end{center}
	\caption{(Colour online) The examples of adsorption-desorption isotherms at different temperatures (given in the figure)
		for the systems with $e=0.6$, $s=0.62$, $\bar{\varepsilon}_{gs}=5$, and $H=10$. Panels a, b, and c correspond to 
		$\Delta V =2.0$, while panels d and e to $\Delta V =2.4$. Circles correspond to the total density, while squares and diamonds 
		correspond to the densities of components A and B, respectively.}
	\label{fig_4}
\end{figure}

Now, we turn to the discussion of systems with selective walls. The first series of calculations were performed for 
mixtures with $e=0.6$, and $s$ ranging between 0.58 and 0.68, adsorbed in pores of~$H$ between 4 and 12, and with 
$\bar{\varepsilon}_{gs}=5$. The main goal of our study is to elucidate the changes in the phase behavior resulting from the wall selectivity, measured by $\Delta V$. 
To this end, we calculated adsorption-desorption isotherms and estimated phase diagrams for several systems. 
Since all the systems studied have shown qualitatively the same changes in phase behavior due to changes in $\Delta V$, 
only the results for selected systems are shown. Figure~\ref{fig_4} presents representative examples
of adsorption-desorption isotherms for the systems  with $s=0.62$, $H=10$, 
and two different values of $\Delta V$ equal to $2.0$ (panels a and~b), and $2.4$ (panels c and d).

\begin{figure}[h!]
	\begin{center}
		\includegraphics[scale=0.4]{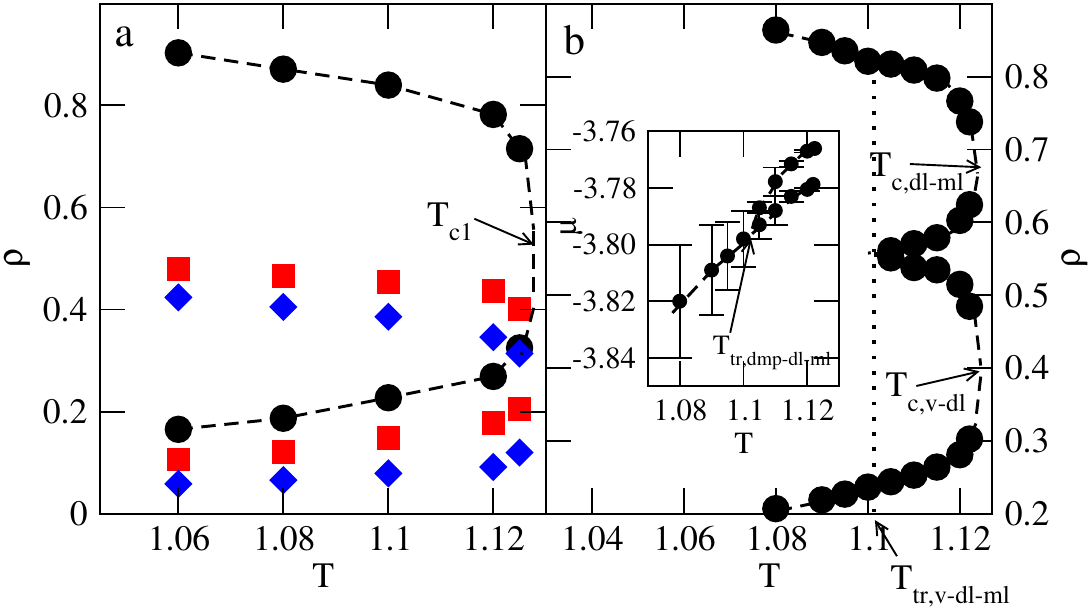}
	\end{center}
	\caption{(Colour online) The $\rho-T$ projections of phase diagrams for the systems characterized by $e=0.6$, $s=0.62$, $\bar{\varepsilon}_{gs}=5$, and $H=10$, for
		$\Delta V =2.0$ (panel a), and $2.4$ (panel b). In panel a, we have shown the densities of the components A (filled squares), and B (filled diamonds), 
		and the total density (filled circles). 
		The inset to panel b shows the estimated $\mu-T$ projection of the phase diagram for $\Delta V =2.4$.}
	\label{fig_5}
\end{figure}

In the case of $\Delta V = 2.0$, the capillary condensation leads to a mixed liquid at any temperature up to the capillary condensation critical point,
$T_{\textrm{c1}}$, and the phase diagram (see figure~\ref{fig_5}a) is qualitatively the same as for one-component systems. Of course, the difference between 
the strengths of interaction between the components and the pore walls causes the density of component A to be higher than the density of component B.

When $\Delta V =2.4$, the phase behavior changes (see panels d and e of figure~\ref{fig_4}). At $T=1.10$, the condensation of a dilute phase leads to a mixed liquid, 
but at $T=1.11$ the dilute phase condenses into the demixed liquid. Only when the chemical potential is sufficiently high,  the demixed liquid undergoes the transition to the mixed liquid of higher
density (cf. panels d and e of figure~\ref{fig_4}). This implies that there is a triple point $T_{\textrm{tr,v-dl-ml}}$, in which the dilute phase coexists with the demixed and mixed 
liquids (see figure~\ref{fig_5}b). 
It should be noted that the transitions between the dilute phase and the demixed liquid, and between the demixed and mixed liquids, 
both terminate in the respective critical points, $T_{\textrm{c,v-dl}}$ and $T_{\textrm{c,dl-ml}}$. Thus,
unlike in the case of systems with $s>s_o$, the continuous demixing transition does not occur at all. 
In the particular system with $e=0.6$, $s=0.62$, $\bar{\varepsilon}_{gs}=5$, and $H = 10$, the critical temperatures $T_{\textrm{c,v-dl}}$ and $T_{\textrm{c,dl-ml}}$ are practically the same.
However, the estimated phase diagrams for several systems characterized by different $s$, $H$, $\bar{\varepsilon}_{gs}$, and $\Delta V$ have shown that 
the critical temperatures $T_{\textrm{c,v-dl}}$ and $T_{\textrm{c,dl-ml}}$ may be slightly different, with $T_{\textrm{c,v-dl}}$ higher or lower than $T_{\textrm{c,dl-ml}}$.

\begin{figure}[h!]
\begin{center}
 \includegraphics[scale=0.5]{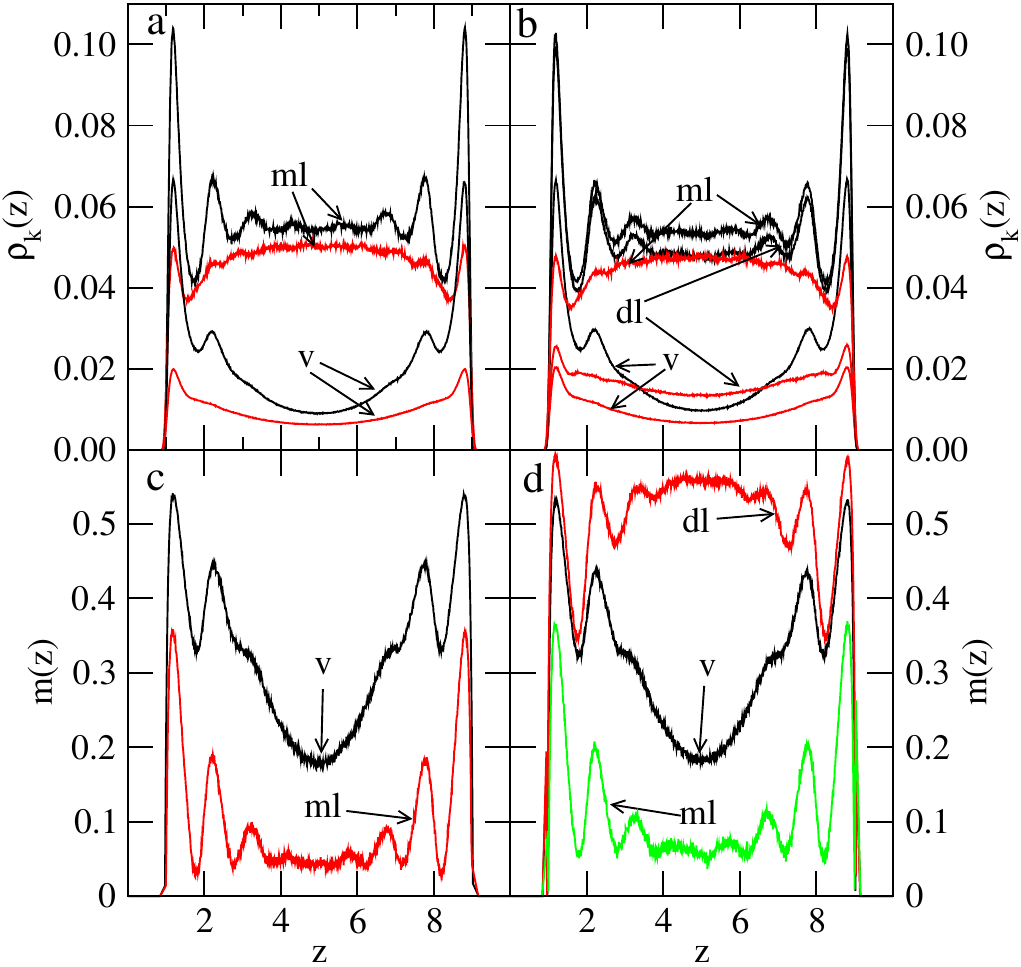}
 \end{center}
 \caption{(Colour online) The density profiles (panels a and b) and the profiles of the order parameter $m(z)$ (panels c and d) for the system with $e=0.6$, $s=0.62$, $\bar{\varepsilon}_{gs}=5$,  
 $\Delta V= 2.4$, and $H=10$, at $T=1.10$ and $\mu=-3.79$ (panels a and c), and at $T=1.11$ and $\mu=-3.788$ (v), $-3.784$ (dl), and $-3.78$ (ml) (panels b and~d).} 
 \label{fig_6} 
\end{figure}

\begin{figure}[h!]
	\begin{center}
		\includegraphics[scale=0.4]{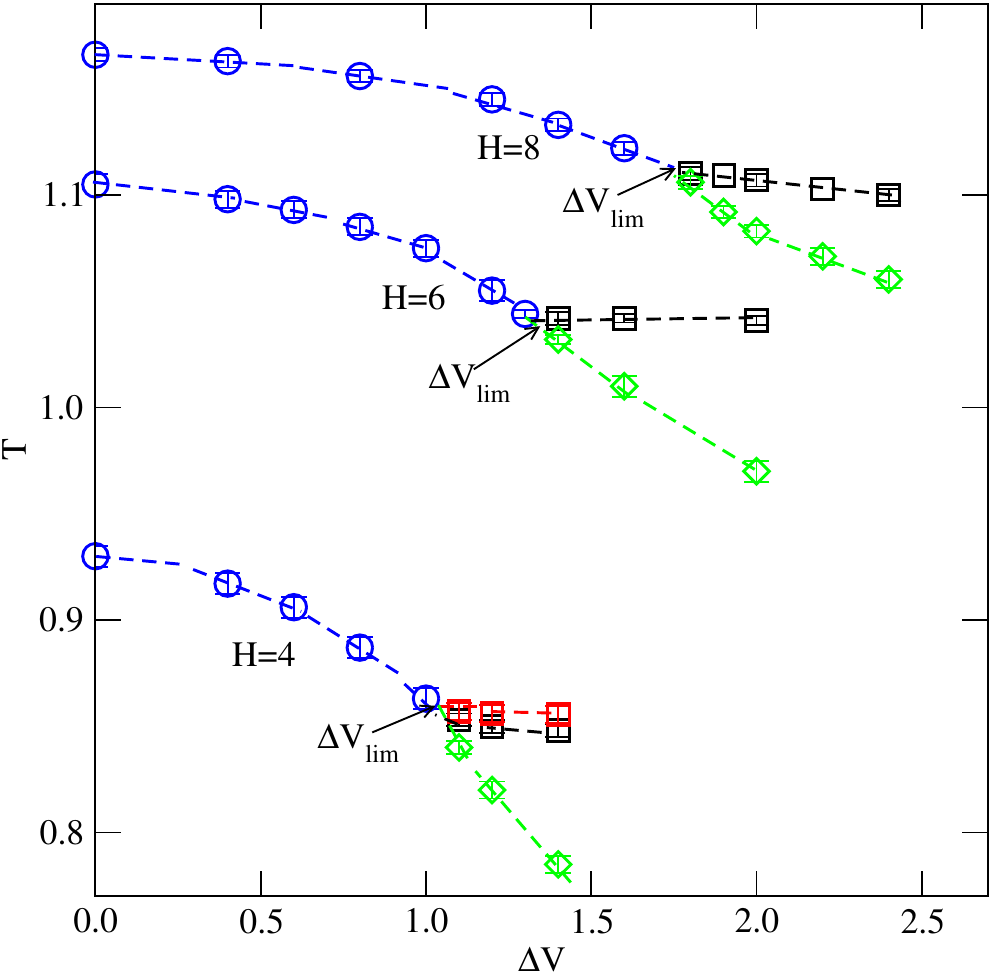}
	\end{center}
	\caption{(Colour online) The changes of $T_{\textrm{c1}}$ (circles), $T_{\textrm{cv-dl}}$ (squares), $T_{\textrm{c,dl-ml}}$ (diamonds), and $T_{\textrm{tr,v-dl-ml}}$ for the system with
		$s=0.58$, $=0.60$, and $\bar{\varepsilon}_{gs}=5$, in pores of different widths (given in the figure).} 
	\label{fig_7} 
\end{figure}

The behavior of systems with $\Delta V=2.0$ and 2.4 makes it clear that there exists a certain limiting value of $\Delta V$, equal to $\Delta V_{\rm lim}$,
such that for any $\Delta V <\Delta V_{\rm lim}$, the demixed liquid does not appear, while for $\Delta V> \Delta V_{\rm lim}$ the first step of capillary condensation  
leads to the formation of demixed liquid, followed by the transition between the demixed and mixed liquids. 
When $\Delta V$ approaches $\Delta V_{\rm lim}$ from above, the triple point $T_{\textrm{tr,v-dl-ml}}$ moves up and terminates in the point where 
the critical temperatures $T_{\textrm{c,v-dl}}$ and $T_{\textrm{c,dl-ml}}$
meet the critical point $T_{\textrm{c1}}$.

In order to demonstrate the changes in the structure of adsorbed fluid in different phases, we present the examples of
the density profiles $\rho_k(z)$,
and the profiles of the order parameter $m(z)$, recorded for the system with $e=0.6$, $s=0.62$, $\bar{\varepsilon}_{gs}=5$, $\Delta V= 2.4$, and $H=10$, 
at $T=1.10$ (panels a and c of figure~\ref{fig_6}), and at $T=1.11$ (panels b and d of figure~\ref{fig_6}). Close to the pore walls, the dilute phase is highly demixed due to a large difference 
in the fluid-wall interaction energies. On the other hand, in the central part of the pore, the difference in the density of the components is lower.
When the capillary condensation leads to the formation of a mixed liquid, the difference between the densities of the components in the central part of the pore is considerably lower
than in the dilute phase. This also occurs in the vicinity of the pore walls, as the  order parameter profiles given in figure~\ref{fig_6}c demonstrate.
When the demixed liquid appears (see panels b and d of figure~\ref{fig_6}), the profile of the order parameter, $m(z)$, takes on quite large values in the central part of the pore. 
A subsequent transition to the mixed liquid causes $m(z)$ to drop and to assume quite low values in the central part of the pore, as well as in the layers 
adjacent to the pore walls. 
Of course, the value of $\Delta V_{\rm lim}$ depends on the parameters $e$ and $s$, the pore diameter, 
as well as on the magnitude of~$\bar{\varepsilon}_{gs}$. In order to demonstrate these effects, we have estimated phase diagrams, 
as those given in figure~\ref{fig_5}, for several systems.
\begin{figure}[h]
\begin{center}
 \includegraphics[scale=0.4]{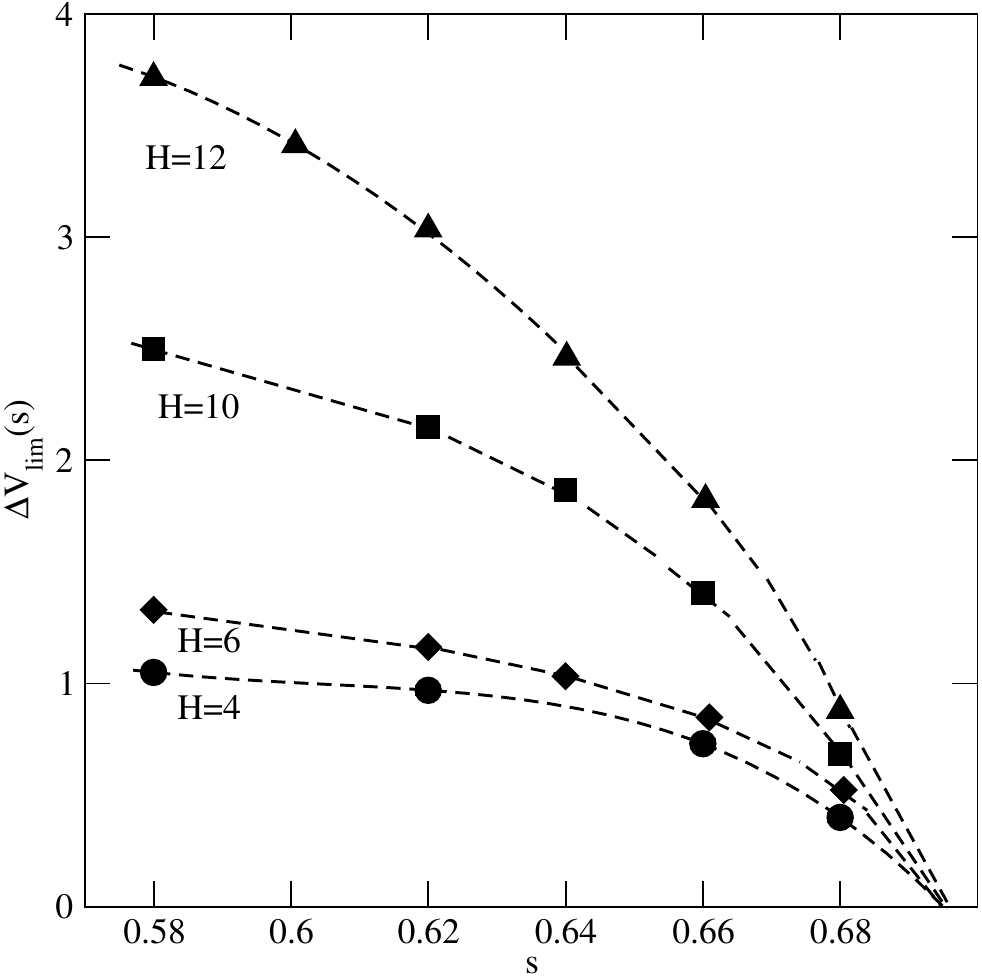}
 \end{center}
  \caption{The estimated changes of $\Delta V_{\rm lim}$ with $s$ for the systems with $e=0.6$ and the pores of different width (given in the figure).}
  \label{fig_8}
 \end{figure}

For the systems with fixed values of $e$, $s$, $H$ and $\bar{\varepsilon}_{gs}$, 
the changes of critical temperatures, $T_{\textrm{c1}}$, $T_{\textrm{c2}}$, $T_{\textrm{c3}}$ and of the triple point temperature $T_{\textrm{tr,v-dl-ml}}$ 
with $\Delta V$, allow us to estimate $\Delta V_{\rm lim}$. 
 The examples of the results obtained for the systems with $s=0.58$, $e=0.6$, and $\bar{\varepsilon}_{gs}=5$, in pores of different widths are shown in figure~\ref{fig_7}. 
One should note that the triple point $T_{\textrm{tr,v-dl-ml}}$ meets the critical 
temperatures $T_{\textrm{c1}}$, $T_{\textrm{c,v-dl}}$ and $T_{\textrm{c,dl-ml}}$ at different values of $\Delta V$. 
The estimated changes of $\Delta V_{\rm lim}$ with $s$ for systems with $e=0.6$, and $\bar{\varepsilon}_{gs}=5.0$, are summarized in figure~\ref{fig_8}.
It is evident that for any $H$, $\Delta V_{\rm lim}$ decreases when $s$ increases, and converges to zero when $s$ reaches the value of about 0.695, i.e., when $s$ becomes equal 
to $s_o$. We should recall that 
in the bulk, as well as in slit pores with non-selective walls, the mixtures with $e=0.6$, and $s$ higher than $s_o$,  the vapor condenses directly into the mixed liquid only at 
the temperatures between the triple points $T_{\textrm{tr,v-dl-ms}}$ and $T_{\textrm{tr,v-dl-ml}}$.
At temperatures above the triple point $T_{\textrm{tr,v-dl-ml}}$, the gas-demixed liquid transition occurs (cf. figure~\ref{fig_1}c and d). 
The same occurs in slit pores with non-selective walls~\cite{AP-JML}, and the vapor-demixed liquid, as well as the demixed liquid-mixed liquid
transitions terminate in the respective tricritical points~\cite{AP-JML}.
The results presented in figure~\ref{fig_8} also demonstrate that for given $e$ and $s$, $\Delta V_{\rm lim}$ quite rapidly increases when the pore becomes wider. 
This can be readily understood, since the increase of $H$ makes the system more and more similar to the bulk, in which the demixed liquid does not appear at all.

\begin{figure}[h]
	\begin{center}
		\includegraphics[scale=0.4]{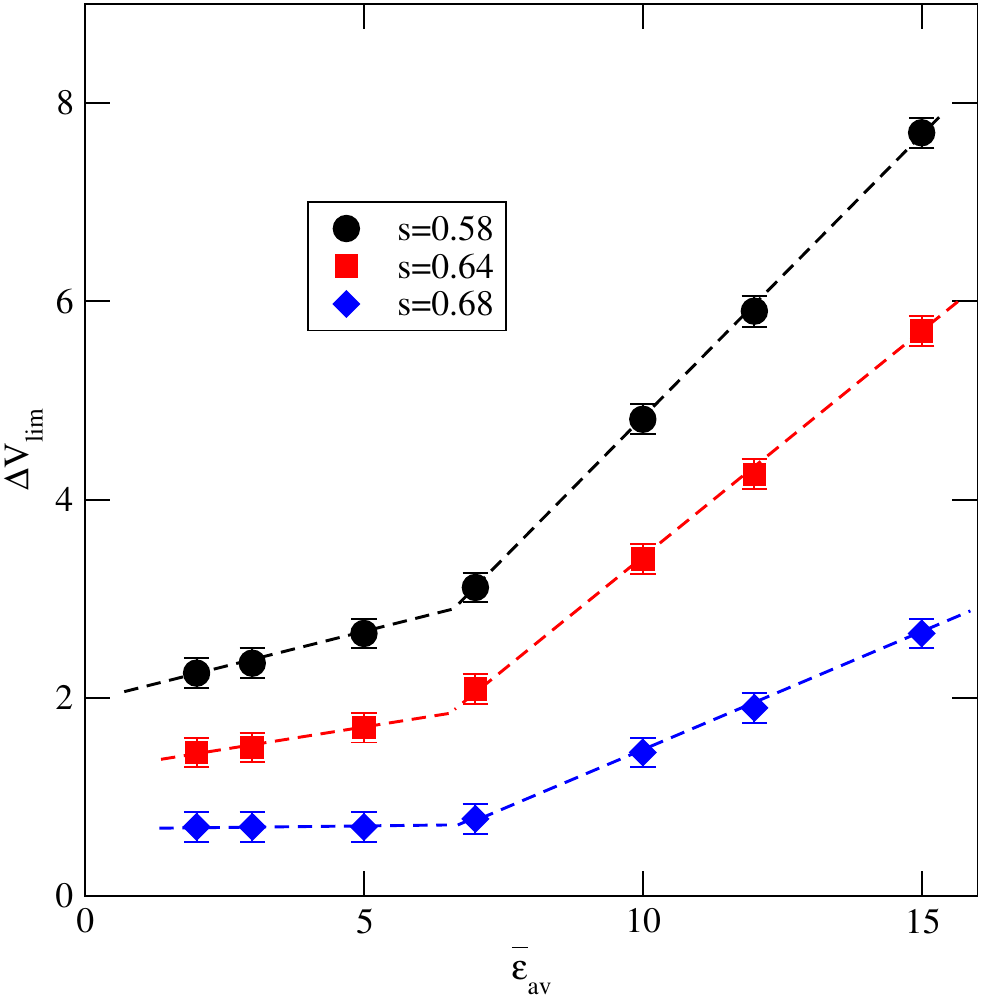}
	\end{center}
	\caption{(Colour online) The changes of $\Delta V_{\rm lim}$ with $\bar{\varepsilon}_{gs}$ for three different values of $s$ (given in the figure), when $e=0.6$ and $H=10$.}
	\label{fig_9}
\end{figure}

Now, we consider the effects of $\bar{\varepsilon}_{gs}$ on $\Delta V_{\rm lim}$, and present explicit results for the systems characterized by $e=0.6$ 
and different values of $s$, equal to 0.58, 0.64, and 0.68 adsorbed in the pore of $H=10$. Figure~\ref{fig_9} shows the estimated changes of
$\Delta V_{\rm lim}$ with $\bar{\varepsilon}_{gs}$. It is quite clear that there are two regions of $\bar{\varepsilon}_{gs}$ over which $\Delta V_{\rm lim}$
exhibits different behavior. For the values of $\bar{\varepsilon}_{gs}$ lower than about 7.0, $\Delta V_{\rm lim}$ only weakly depends on $\bar{\varepsilon}_{gs}$, while
for the larger average strength 
of interaction between the mixture components and the pore walls, $\Delta V_{\rm lim}$ rather rapidly grows with $\bar{\varepsilon}_{gs}$. 

\begin{figure}[h]
	\begin{center}
		\includegraphics[scale=0.4]{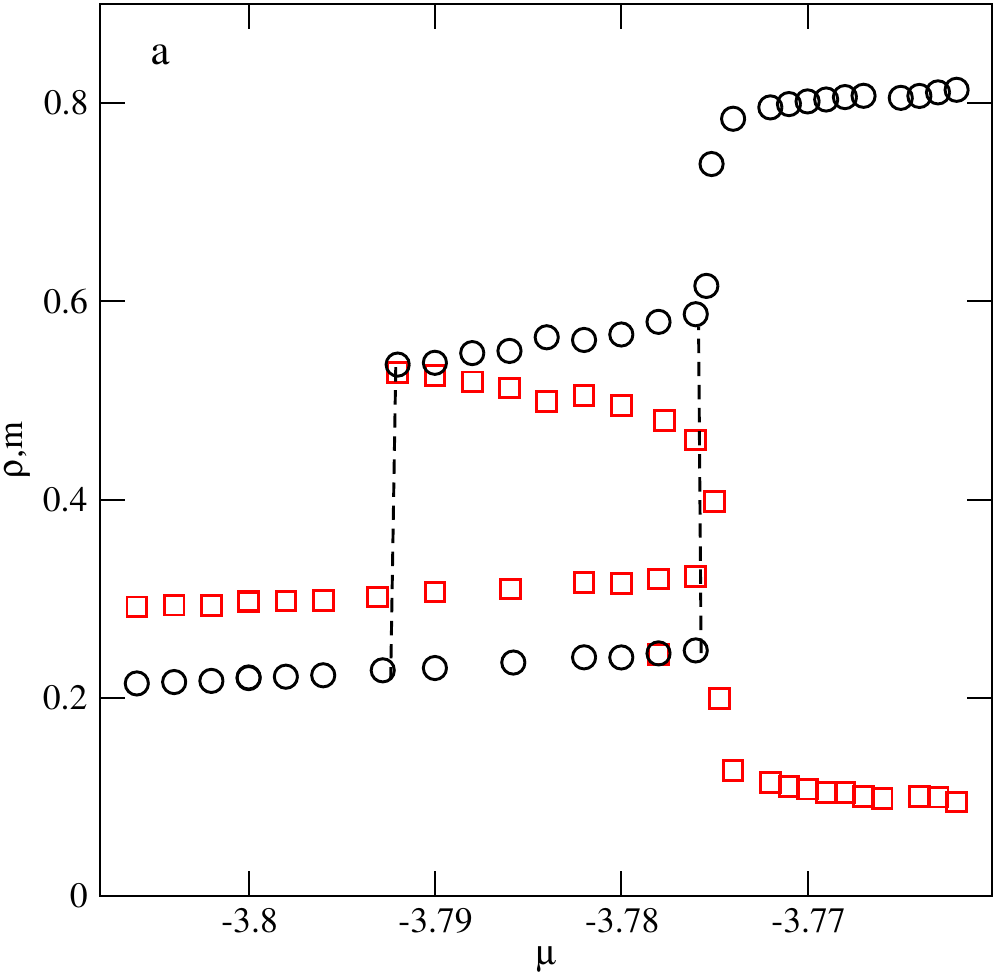}\includegraphics[scale=0.4]{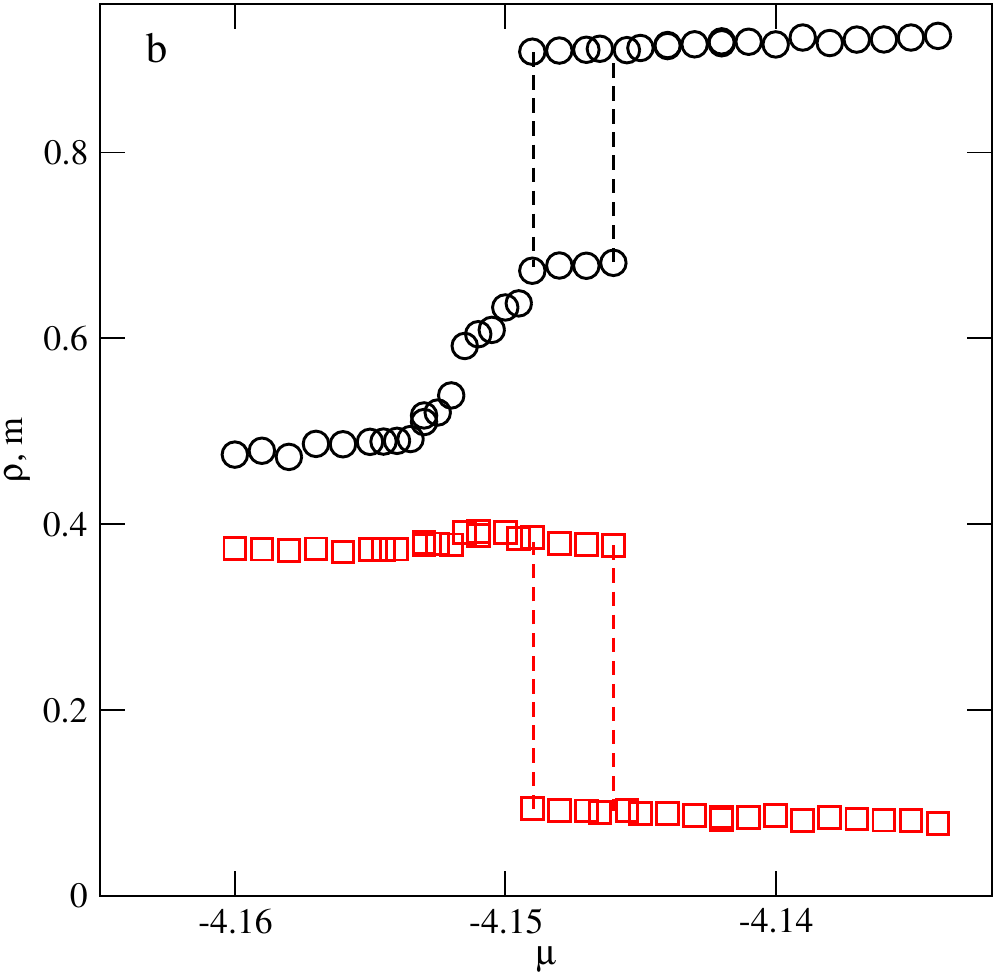}
	\end{center}
	\caption{(Colour online) The examples of isotherms (circles) and the isothermal changes of the order parameter $m$ (squares) for the systems with 
		$e=0.6$, $s=0.64$, and $H=10$, recorded for $\bar{\varepsilon}_{gs} = 5$ and $\Delta V = 2.0$ at 
		$T =1.10$ (panel a) and for $\bar{\varepsilon}_{gs} = 10$ and $\Delta V = 3.7$ at $T=1.06$ (panel b)}.
	\label{fig_10}
\end{figure}

This results from the changes in the structure of low-density phases and the relative stability of demixed and mixed dense phases, 
which leads to different phase behavior of systems with low and high values of $\bar{\varepsilon}_{gs}$. Figure~\ref{fig_10} shows two examples of adsorption-desorption
isotherms for $s=0.64$, obtained for $\bar{\varepsilon}_{gs}$ equal to 5.0 (panel a) and 10.0 (panel b). In both cases, we have used 
$\Delta V > \Delta V_{\rm lim}$, so that the phase diagrams of both systems are expected to be qualitatively the same as that shown in figure~\ref{fig_5}b. 

It is evident that this is true when  
$\bar{\varepsilon}_{gs} = 5$ and $\Delta V = 2.0$.  In this case, the critical temperature $T_{\textrm{c,v-dl}}$ is higher than the critical temperature $T_{\textrm{c,dl-ml}}$.
On the other hand, when $\bar{\varepsilon}_{gs}=10$ and $\Delta V= 3.7$, the formation of demixed liquid occurs gradually, and only the transition between the dl and dm 
phases is discontinuous and terminates at the critical point $T_{\textrm{c,dl-ml}}$. The calculations performed at lower temperatures have never shown the first-order 
condensation of a dilute phase into the demixed liquid. The same results were  obtained for the systems with larger values of 
$\bar{\varepsilon}_{gs}$ equal to 12 and 15, and when $\Delta V > \Delta V_{\rm lim}$. This behavior can be understood by taking into account that the adsorbed layers 
formed at the pore walls become thick enough to suppress the transition to the demixed liquid.   

The results obtained for the systems with $\bar{\varepsilon}_{gs}=7$ and different values of $\Delta V > \Delta V_{\rm lim}$ have shown 
that $T_{\textrm{c,v-dl}}$ is only slightly lower than $T_{\textrm{c,dl-ml}}$. One should also note the difference in the behavior of the order parameter $m$ 
in the case of low and high $\bar{\varepsilon}_{gs}$. When $\bar{\varepsilon}_{gs}= 5$ and $\Delta V = 2.0$, the parameter $m$ in the demixed 
liquid is considerably higher than in the dilute phase. On the other hand, when $\bar{\varepsilon}_{gs}= 10$ and $\Delta V = 3.7$, 
the parameter $m$ is nearly the same in the dilute phase and in the demixed liquid. The order parameter profiles recorded at $\mu=-4.148$ and $-4.16$, shown in the right-hand lower panel 
of figure~\ref{fig_10}, demonstarte a rather large increase of $m(z)$ in the central part of the pore. The density and order parameter profiles also demonstrate that in the case of 
$\bar{\varepsilon}_{gs}= 5$, the adsorbed layers adjacent to the walls are thinner than in the case of $\bar{\varepsilon}_{gs}= 10$, and the thickness of these layers is higher when  
$\bar{\varepsilon}_{gs}$ is larger. Besides, in the system with $\bar{\varepsilon}_{gs}= 10$, the layers adjacent to the walls are considerably denser than in the case of 
$\bar{\varepsilon}_{gs}= 5$. 
We have to emphasize that quite similar results have been obtained for 
other systems with $\bar{\varepsilon}_{gs}$ lower and higher than 7.0, and with $\Delta V > \Delta V_{\rm lim}$. 

\begin{figure}[h]
\begin{center}
\includegraphics[scale=0.4]{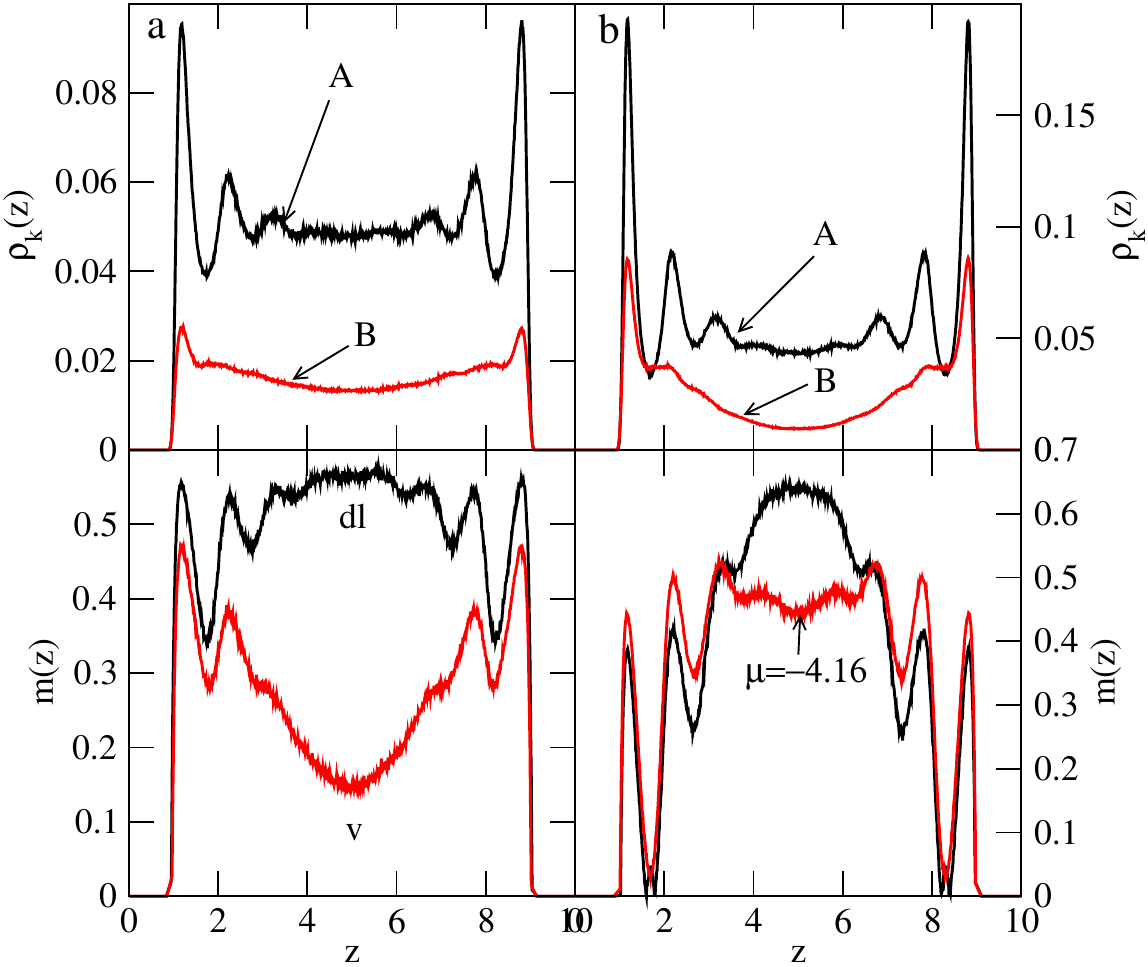}
 \end{center}
 \caption{(Colour online) The density profiles (upper panels) and the corresponding order parameter profiles (lower panels) for the systems with 
 $e=0.6$, $s=0.64$, and $H=10$, recorded for $\bar{\varepsilon}_{gs} = 5$ and $\Delta V = 2.0$ at 
$T =1.10$ and $\mu = -3.786$ (panel a) and for $\bar{\varepsilon}_{gs} = 10$ and $\Delta V = 3.7$ at $T=1.06$ and $\mu = 4.148$ (panel b).
In the lower panel of panel b, we have shown the order parameter profile at the chemical potential $\mu=-4.16$, well below the formation of the 
demixed liquid.}
\label{fig_11}
\end{figure}

Figure~\ref{fig_11} presents the density and order parameter profiles recorded at certain points along the isotherms shown in figure~\ref{fig_10}.
It is well seen that in the demixed liquid, the densities of the components in the first two layers close to the pore walls are lower when $\bar{\varepsilon}_{gs}=5$ and $\Delta V = 2.0$, and considerably higher when $\bar{\varepsilon}_{gs}=10$ and $\Delta V = 3.7$.
On the other hand, the density of component A in the central part of the pore is very similar in both systems, whereas the density of component B in the central part of the pore is higher than in the system with  weakly adsorbing walls. This results in lower values of the order parameter in the central part of 
the pore, in the system with $\bar{\varepsilon}_{gs}=5$ and $\Delta V = 2.0$ than in the case of $\bar{\varepsilon}_{gs}=10$ and $\Delta V = 3.7$ (see lower panels 
of figure~\ref{fig_11}). The lower panel of figure~\ref{fig_11}b shows the order parameter profile recorded at the chemical potential well below the formation of demixed liquid, and
demonstrates that the surface film has practically the same thickness, and only in the central part of the pore, the order parameter reaches higher values in the 
demixed liquid. 

\begin{figure}[h]
\begin{center}
\includegraphics[scale=0.3]{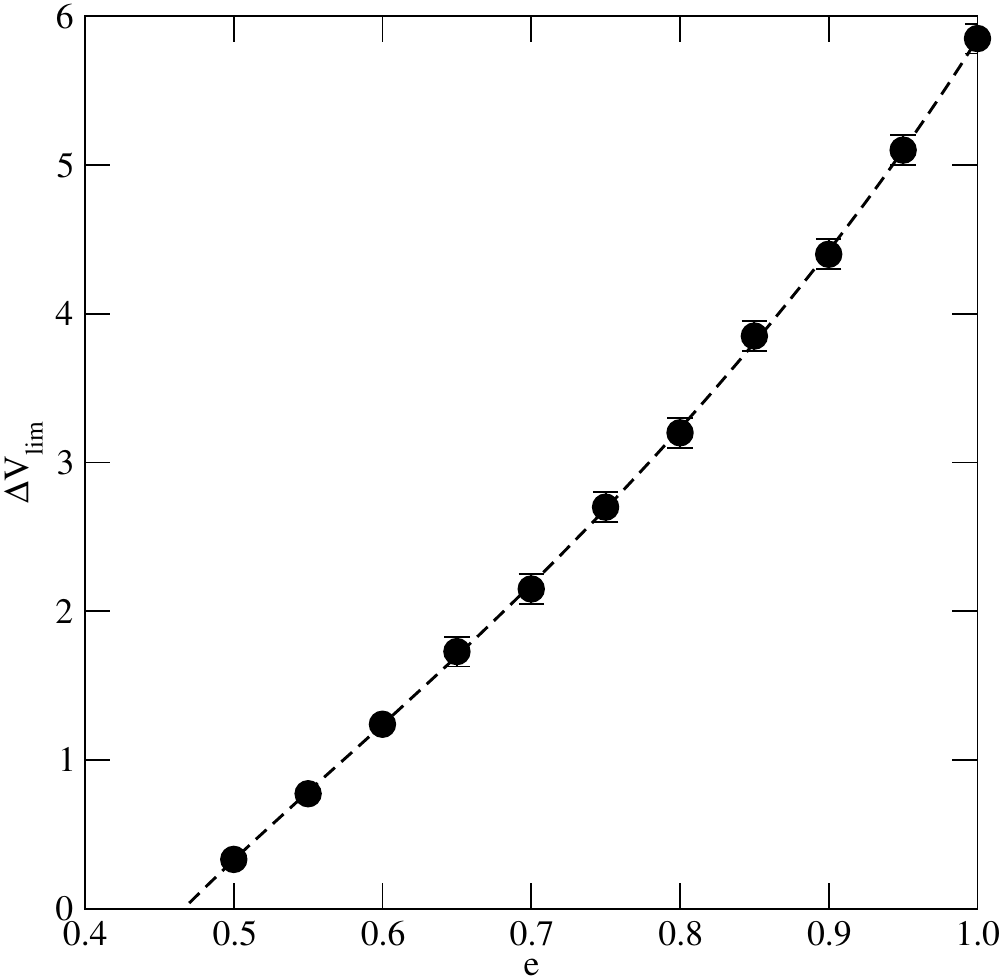}
\end{center}
\caption{The changes of $\Delta V_{\rm lim}$ with $e$ for the systems with $s=0.6$, $\bar{\varepsilon}_{gs}=5.0$, and $H=6$.}  
\label{fig_12}
\end{figure}

The last question we wanted to address in this work concerns  
the influence of packing effects  on the demixing transition in the pores with selective walls. We have considered a series of systems with the fixed values of $s=0.6$, 
$\bar{\varepsilon}_{gs}=5.0$, and $H=6$, while the parameter $e$ was varied. The estimated changes of $\Delta V_{\rm lim}$ with $e$ are shown in figure~\ref{fig_12}, 
and demonstrate that the demixing occurs even for quite large values of $e$. It should be remembered that an increase of $e$ reduces the tendency towards demixing, leading to 
a gradual increase of $\Delta V_{\rm lim}$. Of course, a decrease of $e$ causes that $\Delta V_{\rm lim}$ also decreases and tends to zero for 
$e\approx0.46$. 

We have estimated the phase diagrams for bulk systems with $s=0.6$ and different values of $e$, which have demonstrated that 
the demixing transition occurs for any $s$ lower than about 0.48. A characteristic feature of the vapor-liquid (mixed or demixed) condensation in bulk 
symmetric mixtures with high geometric and energetic non-additivity is a negative slope of   
the chemical potential  with respect to temperature. Since it is equal to the negative of the specific entropy of the phase, the negative slope 
of the chemical potential at gas-liquid condensation represents the release of entropy during the phase change from a higher-entropy gas to a lower-entropy liquid. 
 In the system with $s=0.6$ and $s=0.5$, which does not show the demixing transition, the slope of the chemical potential along the gas-liquid coexistence is positive, 
 just the same as in one-component systems. 

\section{Summary}
\label{sec-4}

The paper presents the results of Monte Carlo simulations for symmetric mixtures in slit pores, which, due to high energetic and geometric non-additivity, 
do not exhibit phase separation in bulk systems. Such mixtures also do not exhibit phase separation in slit pores when both components interact equally 
with the pore walls, i.e., in the case of non-selective walls. Therefore, capillary condensation leads to the formation of a mixed liquid phase, and the 
critical temperatures of capillary condensation vary with pore width according to the scaling relation described by equation (\ref{eq-scala}). Of course, this relation 
is only satisfied when the pore widths are sufficiently large. We have shown that despite only an approximate estimation of critical temperatures, 
relation (\ref{eq-scala}) is satisfied for pores with a width of $H\geqslant 10$. It should be mentioned that Nakanishi and Fisher~\cite{Fish01} suggested that the scaling relation (\ref{eq-scala}) 
should hold when the pore diameter is greater than about $5\sigma$. 

The influence of pore wall selectivity on capillary condensation was investigated. It was shown that for pores of a given width, $H$ the mechanism of capillary condensation depends on the difference in the interaction energy between the mixture components and the pore walls ($\Delta V$). When this difference is less than a certain limit value, $\Delta V_{\rm lim}$, 
capillary condensation leads to the formation of a mixed liquid phase (ml).  When, however, $\Delta V > \Delta V_{\rm lim}$, and the temperature is sufficiently low,
capillary condensation proceeds in two stages. 
The first stage leads to a demixed liquid phase (dl), and the second stage is associated with the formation of a denser, mixed liquid phase. At low temperatures, 
both stages of capillary condensation occur as phase transitions of the first order, ending at respective critical points. Therefore, when $\Delta V > \Delta V_{\rm lim}$, 
a triple point occurs where the gas phase coexists with the mixed and demixed liquid phases (cf. figure~\ref{fig_7}).  
$\Delta V_{\rm lim}$ depends on $H$, $s$, $e$, and $\bar{\varepsilon}_{gs}$. As expected,  $\Delta V_{\rm lim}$ gradually increases with increasing pore width. 
An increase in the $s$ parameter also increases the tendency for phase separation, and thus leads to a gradual decrease in $\Delta V_{\rm lim}$. It has been shown that, 
regardless of pore width, $\Delta V_{\rm lim}$ tends toward zero when the $s$ parameter reaches a value above which phase separation occurs in bulk systems (cf. figure~\ref{fig_8}). 

We have also demonstrated that the magnitude of $\bar{\varepsilon}_{gs}$ influences the phase behavior considerably. For $\bar{\varepsilon}_{gs}$ lower than about 7.0, the 
critical temperature $T_{\textrm{c,v-dl}}$ is higher than $T_{\textrm{c,dl-dm}}$, while for $\bar{\varepsilon}_{gs}$ higher than 7.0 the situation is different.
At the temperatures used, we have always observed a gradual development of the demixed liquid, and the first- order transition between the demixed and mixed 
liquids at the tempartures lower than $T_{\textrm{c,dl-dm}}$. Of course, it is possible that the transition between the dilute gas-like and demixed liquid phases is discontinuous at 
still lower temperatures. If this is the case, then the critical temperature $T_{\textrm{c,v-dl}}$ is lower than the critical temperature $T_{\textrm{c,dl-dm}}$. 

In summary, we have shown that symmetric mixtures, which do not exhibit phase separation in slit pores with non-selective walls, can undergo phase 
separation when the interactions of the mixture components with the pore walls are different. This is due to the changes in the stability of the mixed and unmixed liquid phases.

\ukrainianpart

\title{Симетричні суміші в щілиноподібних порах із селективними стінками}
%
%
\author{A. Патрикеєв}

\address{Кафедра теоретичної хімії, Інститут хімічних наук, Хімічний факультет, Університет Марії Склодовської-Кюрі, 20031 Люблін, Польща}

\makeukrtitle

\begin{abstract}
	\tolerance=3000%
	Симетричні суміші, що характеризуються високою від'ємною геометричною та енергетичною неадитивністю, не демонструють фазового розшарування в об'ємі. Однак фазове розділення відбувається, коли такі суміші утримуються в щілинних порах із селективними стінками. Показано, що селективність стінок впливає на заповнення пор. Коли різниця енергій взаємодії між компонентами суміші та стінками пор нижча за певне порогове значення, відбувається конденсація між розведеною фазою та змішаною рідиною. Коли ця різниця перевищує порогове значення, заповнення пор може відбуватися у два етапи. Перший --- це конденсація розведеної фази у розведену рідину, а другий етап призводить до утворення змішаної рідини. Ми з'ясували зміни у фазовій поведінці, спричинені неадитивністю симетричних сумішей, та різницею в енергіях взаємодії компонент зі стінками пор.
	\keywords симетричні суміші, щілинні пори з селективними стінками, розшарування, моделювання методом Монте-Карло
	
\end{abstract}

\lastpage
\end{document}